\RequirePackage{fixltx2e}
\documentclass[aip,jcp,reprint,floatfix]{revtex4-1}

\usepackage{amsmath,amsfonts} %
\usepackage{mathrsfs} %
\usepackage{wasysym} %
\usepackage[acronym]{glossaries} %
\usepackage{graphicx} %
\usepackage{dcolumn} %
\usepackage{bm} %
\usepackage{ifpdf} %
\usepackage{color} %
\usepackage{nicefrac} %
\usepackage{natbib} %
\usepackage[byname]{smartref} %
\usepackage[breaklinks, %
colorlinks, %
linkcolor=blue, %
citecolor=blue, %
urlcolor=blue]{hyperref} %
\usepackage[table]{xcolor} %

\DeclareMathOperator{\tr}{Tr} %
\newcommand{\eq}[1]{Eq.~(\ref{#1})} %
\newcommand{\eqs}[1]{Eqs.~(\ref{#1})} %
\newcommand{\bea}{\begin{eqnarray}}
\newcommand{\eea}{\end{eqnarray}}

\newcommand{\ket}[1]{\ensuremath{\left|#1\right\rangle}}

\newcommand{\bra}[1]{\ensuremath{\left\langle #1\right|}}

\newcommand{\braket}[2]{\ensuremath{\left\langle #1|#2\right\rangle}}

\newcommand{\braOket}[3]{\ensuremath{\left\langle #1\left|#2\right|#3\right\rangle}}

\newcommand{\op}[1]{\ensuremath{\hat{#1}}}

\newcommand{\mat}[1]{\ensuremath{\boldsymbol{#1}}}
\newcommand{\dt}[1]{\ensuremath{\overset{\bm .}{#1}}}

\newcommand{\inumb}[0]{\dot{\imath}}

\newcommand{\Liouv}[0]{\mathcal{L}}
\newcommand{\sqrtLiouv}[0]{\mathcal{K}}

\newcommand{\Lag}[0]{\Lambda}
\newcommand{\lag}[0]{\lambda}

\renewcommand{\Re}{\operatorname{Re}}
\renewcommand{\Im}{\operatorname{Im}}

\ifpdf %
 \fi
                            
\newacronym{CI}{CI}{conical intersection} %
\newacronym{LVC}{LVC}{linear vibronic coupling} %
\newacronym[firstplural={degrees of freedom (DOF)},plural={DOF},longplural={degrees of freedom}]{DOF}{DOF}{degree of freedom} %
\newacronym{PES}{PES}{potential energy surface} %
\newacronym{BMA}{BMA}{bis(methylene) adamantyl} %
\newacronym{FC}{FC}{Franck-Condon} %
\newacronym[firstplural={density matrices (DMs)},plural={DMs},longplural={density matrices}]{DM}{DM}{density matrix} %
\newacronym{LvN}{LvN}{Liouville-von Neumann} %
\newacronym{QME}{QME}{quantum master equation} %
\newacronym{TDSE}{TDSE}{time-dependent Schr\"odinger equation} %
\newacronym{NOSSE}{NOSSE}{non-stochastic open system Schr\"odinger equation} %
\newacronym{TDVP}{TDVP}{time-dependent variational principle} %
\newacronym{vMCG}{vMCG}{variational multiconfiguration Gaussian} %
\newacronym{MCTDH}{MCTDH}{multiconfiguration time-dependent Hartree} %
\newacronym{MLMCTDH}{MLMCTDH}{multi-layer MCTDH} 
\newacronym{AIMS}{AIMS}{ab initio multiple spawning} %
\newacronym[firstplural={equations of motion (EOM)},plural={EOM},longplural={equations of motion}]{EOM}{EOM}{equation of motion} %
\newacronym{CS}{CS}{coherent state} %
\newacronym{RMSD}{RMSD}{root-mean-square deviation} %

\begin{document}

\title{Problem-free time-dependent variational principle for open quantum systems}

\author{Lo{\"i}c Joubert-Doriol} %
\affiliation{Department of Physical and Environmental Sciences,
  University of Toronto Scarborough, Toronto, Ontario, M1C 1A4,
  Canada} %
\affiliation{Chemical Physics Theory Group, Department of Chemistry,
  University of Toronto, Toronto, Ontario M5S 3H6, Canada} %
\author{Artur F. Izmaylov} %
\affiliation{Department of Physical and Environmental Sciences,
  University of Toronto Scarborough, Toronto, Ontario, M1C 1A4,
  Canada} %
\affiliation{Chemical Physics Theory Group, Department of Chemistry,
  University of Toronto, Toronto, Ontario M5S 3H6, Canada} %

\date{\today}

\begin{abstract}
Methods of quantum nuclear wave-function dynamics have become very efficient in simulating large isolated systems  
using the \gls{TDVP}.  However, a straightforward extension of the \gls{TDVP} to the density matrix framework gives rise 
to methods that do not conserve the energy in the isolated system limit and the total system population for open systems
where only energy exchange with environment is allowed. 
These problems arise when the system density is in a mixed state and is simulated using an incomplete basis. 
Thus, the basis set incompleteness, which is inevitable in practical calculations, creates artificial channels for 
energy and population dissipation. To overcome this unphysical behavior, we have introduced a constrained 
Lagrangian formulation of \gls{TDVP} applied to a non-stochastic open system Schr\"odinger equation (NOSSE)
[L. Joubert-Doriol, I. G. Ryabinkin, and A. F. Izmaylov, J. Chem. Phys. \textbf{141}, 234112 (2014)].
While our formulation can be applied to any variational ansatz for the system density matrix, 
derivation of working equations and numerical assessment are done within the \acrlong{vMCG} 
approach for a two-dimensional \acrlong{LVC} model system interacting with a harmonic bath.
\end{abstract}

\pacs{}

\maketitle

\glsresetall

\section{Introduction}
\label{sec:introduction}

Modelling photo-induced quantum dynamics in large molecular systems 
requires consideration of relatively localized chromophoric regions where 
this dynamics originates as well as macroscopic environment 
embedding these regions and influencing their behavior.\cite{Collini:2010/N/644,Robb:2009/JACS/13580,Kukura:2005/S/1006}
This is a very challenging task due to exponential scaling of quantum mechanics with 
the number of \glspl{DOF}.

One successful approach to reduce the prefactor of the exponential scaling for isolated systems 
in pure states is to project the \gls{TDSE} onto a time-dependent basis set. The basis time-dependence 
gives flexibility to minimize an error in an approximate solution of the 
\gls{TDSE} with reduced computational cost as compared to that for static bases.
\Glspl{EOM} for the basis set parameters are obtained using the 
\gls{TDVP}.~\cite{Dirac:1958/Book,Frenkel:1934/Book,Mclachlan:1964/mp/39}
This approach led to methods that are able to describe up to hundreds of nuclear \glspl{DOF} 
such as \gls{MCTDH} single-~\cite{Beck:2000/pr/1,Burghardt:1999/jcp/2927} and multi-layer~\cite{Wang:2003/jcp/1289} methods, 
as well as \gls{vMCG} method.\cite{Worth:2004/fd/307,Izmaylov:2013/jcp/104115}
However, computational costs of these approaches still scale exponentially with the number of \gls{DOF}
so that molecular systems such as proteins (several thousands of \glspl{DOF}) cannot be simulated yet. 
Moreover, wave-function methods cannot describe 
energy dissipation and quantum decoherence of a system interacting with environment in a 
mixed state (e.g., heat reservoir or incoherent sun light\cite{Tscherbul:2014/JPCA/3100}) 
without departing from the \gls{TDSE}.~\cite{Book/Breuer:2002}

The \gls{QME} formalism involving the system \gls{DM} can model interaction of the 
quantum system with large macroscopic environment and describe
associated processes of energy exchange and quantum decoherence.~\cite{Book/Breuer:2002,Nitzan:2006} 
These processes give rise to mixed states of the quantum system, which 
cannot be described with a single wave-function. 
Therefore, the use of a \gls{DM}, $\op\rho(t)$, becomes indispensable, 
and the \gls{TDSE} is replaced by the \gls{QME}
\bea\label{eq:QME}
\dt{\op\rho}(t) & = & \Liouv[\op\rho(t)],
\eea
where the dot stands for the partial time derivative, 
and $\Liouv$ is the Liouvillian super-operator. 
In the case of an isolated system with the Hamiltonian $\op H$, 
\bea
\Liouv[\op\rho(t)]=-\inumb[\op H,\op\rho(t)],
\eea
and \eq{eq:QME} becomes the \acrlong{LvN} equation that generates a unitary evolution 
of the \gls{DM} in which the system energy is conserved.
For an open system, $\Liouv$ also contains a non-unitary part 
that is responsible for energy dissipation and quantum decoherence. 
This non-unitary part is obtained either from purely mathematical reasoning to preserve semi-positivity 
of the system density (e.g., Lindblad approach~\cite{Lindblad:1976/cmp/119}) or derived using perturbation theory with respect to the system-bath coupling, 
Redfield~\cite{Redfield:1965/amr/1}, Caldeira-Legett~\cite{Caldeira:1983/pa/587}, and time-convolutionless formulations.
All these formulations reduce environmental effects to few terms in $\Liouv$ that contain only the system 
\gls{DOF}, hence, excluding environmental \glspl{DOF} from the explicit consideration. This reduction
employs perturbative consideration of system-environment interaction 
and thus is adequate only for weak system-environment couplings.

Practical necessity to have small system-bath couplings usually requires to consider a 
large number of system \glspl{DOF} in the \gls{QME} framework, and thus, it increases 
the computational cost of simulation.
Therefore, use of a time-dependent basis guided by the \gls{TDVP} for simulating system dynamics 
in the \gls{QME} approach seems as an ideal choice for modelling large open systems.
Combining the \gls{TDVP} with \gls{QME} has been attempted in the past,
~\cite{Gerdts:1997/jcp/3017,Raab:1999/jcp/8759,Burghardt:1999/jcp/2927} however, these attempts revealed that resulting 
dynamical equations exhibit unphysical behavior which was traced to basis set incompleteness.  
The first problem is non-conservation of the \gls{DM} 
trace, $\tr\{\dt{\op\rho}(t)\} \neq 0$, for open systems where only the energy but not the matter exchange is allowed.~\cite{Gerdts:1997/jcp/3017,Raab:2000/tca/358} 
The second problem is violation of energy conservation for an isolated 
system in a mixed state,~\cite{Mclachlan:1964/mp/39,Heller:1976/jcp/63,Raab:2000/tca/358} $\tr\{\dot{\op\rho}(t)\op H\} \neq 0$.
It is important to note that in the context of the McLachlan \gls{TDVP} for wave-functions, to conserve the energy,
there is a requirement for a wave-function to be analytic with respect to variational parameters. 
This requirement makes the McLachlan \gls{TDVP}~\cite{Mclachlan:1964/mp/39} 
equivalent to a principle of least action~\cite{Broeckhove:1988/cpl/547} that guarantees the energy 
conservation.~\cite{Kan:1981/pra/2831} However, violation of the energy conservation in the density 
formalism has a different origin and takes place even if the density parameterization is analytic. 
It appears as a result of the \gls{DM} entering quadratically in 
the \gls{TDVP}, which makes $\tr\{\op\rho(t)^2\op H\}$ a conserved quantity~\cite{Raab:2000/tca/358} rather than
the energy $E=\tr\{\op\rho(t)\op H\}$. 

These two problems create artificial channels of population and energy flows that can lead to inadequate results.
A straightforward solution involving corresponding constraints imposed 
through the Lagrange multiplier method~\cite{Heller:1976/jcp/63} 
does not preserve a unitary character of quantum dynamics in the isolated system limit.~\cite{note:supp} 
This can lead to more subtle 
artifacts such as non-conservation the density matrix purity ($\tr\{\op \rho^2\}$). Another, so-called linear mean-field 
approach has been suggested by Raab {\it et al.}~\cite{Raab:1999/jcp/8759} within the \gls{MCTDH} framework to resolve these issues.
However, this method becomes non-variational for open systems. 

The aim of this paper is to resolve the non-conservation problems using a hybrid approach: 
To address the energy non-conservation problem in isolated systems we apply the \gls{TDVP} 
not to the \gls{QME} equation but to its recently developed equivalent, \gls{NOSSE}~\cite{Joubert:2014/jcp/234112}.
\Gls{NOSSE} does not have issues with energy conservation within the \gls{TDVP} framework because 
it is formulated with respect to a square root of the density. The root square dynamics is unitary 
for an isolated system and allows for the full reconstruction of the system density matrix evolution 
by evaluating the square at each time. The trace non-conservation occurs only for open systems 
in \gls{TDVP} \gls{NOSSE} formulation and is resolved by imposing 
conservation of the trace of the density matrix through a Lagrangian multiplier. This constraint does 
not interfere with the unitarity of dynamics in the isolated system limit and thus does not introduce any artifacts.

The rest of the paper is organized as follows. Section \ref{sec:theory} is devoted to exposing
 the non-conservation problems at the formal level and details our solutions. Section \ref{sec:results} 
 provides numerical examples illustrating performance of our \gls{TDVP} on a model system. 
 Section \ref{sec:conclusion} concludes this paper and gives an overview of future work.

\section{Theory}
\label{sec:theory}

\subsection{\protect\gls{TDVP} problems in the density matrix formalism}

We start by illustrating formally origins of the \gls{TDVP} problems in the density matrix formalism. 
We introduce a finite linear subspace $\mathscr{S}$ of the total Hilbert space that contains 
time-dependent, not necessarily orthonormal, basis functions $\{\ket{\varphi_k(t)};k=1,\hdots,N_b\}$.
A projector on $\mathscr{S}$ is built as $\op P_{N_b}(t)=\sum_{kl}\ket{\varphi_k(t)}[\mat S^{-1}(t)]_{kl}\bra{\varphi_l(t)}$, 
where $S_{kl}(t)=\braket{\varphi_k(t)}{\varphi_l(t)}$ is an element of the overlap matrix. 
The time dependency of the $\ket{\varphi_k(t)}$ is given through sets of parameters 
$z_{\alpha k}(t)$, where the Greek subscript labels different parameters 
corresponding to the same basis function.
Our only requirement on the parameterization of the basis is the analyticity condition~\cite{Broeckhove:1988/cpl/547}
\bea\label{eq:analyticity}
\frac{\partial\ket{\varphi_k(t)}}{\partial z_{\alpha k}^*(t)}=0,
\eea
so that the parameterization preserves the energy 
in the limiting case of an isolated system in a pure state. 

In the time-dependent basis, $\op\rho(t)$ is expressed as
\bea\label{eq:ansatz-rho}
\op\rho(t) & = & \sum_{kl}\ket{\varphi_k(t)}B_{kl}(t)\bra{\varphi_l(t)},
\eea
where $B_{lk}(t)$ are time-dependent, Hermitian [$B_{kl}(t)=B_{lk}(t)^*$] coefficients.
For notational simplicity, when it is not essential, the time argument will be skipped in the further consideration.
A straightforward extension of the McLachlan \gls{TDVP} to the density formalism involves 
minimization of the Frobenius norm of the error 
\bea
||\dot{\op\rho}-\Liouv[\op\rho]||=\tr\{(\dot{\op\rho}-\Liouv[\op\rho])^\dagger(\dot{\op\rho}-\Liouv[\op\rho])\}^{1/2},
\eea
which is equivalent to satisfying a stationary condition~\cite{Raab:2000/tca/358}
\bea\label{eq:TDVP-QME}
&\tr\left\{\delta\op\rho\left(\dt{\op\rho} - \Liouv[\op\rho]\right)\right\} = 0.
\eea
Replacing $\op\rho$ by \eq{eq:ansatz-rho} in \eq{eq:TDVP-QME} we obtain 
\gls{EOM} for $\mat B=\{B_{kl}\}$ and $z_{\alpha k}$ (detailed derivation is given in App.~\ref{sec:TDVP-QME})
\bea
& \dt{\mat B} = \mat S^{-1}\mat L\mat S^{-1} - \left(\mat S^{-1}\mat\tau\mat B + \mat B\mat\tau^\dagger\mat S^{-1}\right) &, \label{eq:vMCG-DM-B}\\
& \sum_{\beta l} {\tilde C}^{\alpha\beta}_{kl} {\dt z}_{\beta l} = {\tilde Y}^\alpha_k, & \label{eq:vMCG-DM-lambda1}
\eea
where $[\mat \tau]_{kl}=\braket{\varphi_{k}}{\dot{\varphi}_l}$, $[\mat L]_{kl}=\braOket{\varphi_{k}}{\Liouv[\op\rho]}{\varphi_l}$, and 
\bea\label{eq:vMCG-DM-C}
{\tilde C}^{\alpha\beta}_{kl} & = & \braOket{ \frac{\partial \varphi_k}{\partial  z_{\alpha k}} }
                                         { \left[\op 1 - \op P_{N_b} \right] }
                                         { \frac{\partial \varphi_l}{\partial  z_{l\beta}} } [\mat B\mat S\mat B]_{lk}, \\
{\tilde Y}^\alpha_k & = & \sum_l\braOket{ \frac{\partial \varphi_k}{\partial  z_{\alpha k}} }{
                  \left[\op 1 - \op P_{N_b}\right]\Liouv[\op\rho]}{\varphi_l}B_{lk}.\label{eq:vMCG-DM-Y}
\eea
Equation (\ref{eq:vMCG-DM-lambda1}) is solved for $\dot{\mat z}=\{{\dot z}_{\alpha k}\}$ as a linear equation 
where $\dot{\mat z}$ and $\tilde{\mat Y}$ are vectors and $\tilde{\mat C}$ is a matrix
\bea
\dt{\mat  z} & = & \tilde{\mat C}^{-1}\tilde{\mat Y}. \label{eq:vMCG-DM-lambda}
\eea
A given parameterization can lead to redundancies between some of the parameters $z_{\alpha k}$ and $B_{kl}$ 
(e.g., in the \gls{MCTDH} and \gls{vMCG} methods). In such cases extra relations between redundant 
parameters must be combined with Eqs.~(\ref{eq:vMCG-DM-B}) and (\ref{eq:vMCG-DM-lambda}) to eliminate any redundancies.

Equations~(\ref{eq:vMCG-DM-B}) and (\ref{eq:vMCG-DM-lambda}) are equivalent to \glspl{EOM} obtained 
in the particular case of moving Gaussians derived in Ref.~\onlinecite{Burghardt:1999/jcp/2927}. 
This is not fortuitous since the ``type II'' density parameterization of Ref.~\onlinecite{Burghardt:1999/jcp/2927} 
coincides with that of \eq{eq:ansatz-rho}.
By choosing the \gls{MCTDH} ansatz for the time-dependent basis, it can be verified that 
Eqs.~(\ref{eq:vMCG-DM-B})-(\ref{eq:vMCG-DM-lambda}) provide the \glspl{EOM} 
obtained in Refs.~\onlinecite{Raab:1999/jcp/8759,Raab:2000/tca/358} (up to the removal of redundancies in parameterization).

To illustrate variation of the total system population we will use the derived \gls{EOM} to consider 
time derivative of the system population
\bea\label{eq:not-conserved-trace-QME}
\tr\{\dt{\op\rho}(t)\} =  \tr\{\dt{\mat B}\mat S+(\mat\tau+\mat\tau^\dagger)\mat B\}.
\eea
Using \eq{eq:vMCG-DM-B} the variation of the \gls{DM} trace on time is given by
\bea\nonumber
 \tr\{\dt{\op\rho}(t)\} & = & \tr\{\mat S^{-1}\mat L - \mat S^{-1}\mat\tau\mat B\mat S - \mat B\mat\tau^\dagger+(\mat\tau+\mat\tau^\dagger)\mat B\}\\\label{eq:noTd}
 &=& \tr\{\mat S^{-1}\mat L\} = \tr\{\op P_{N_b}\Liouv[\op\rho]\},
\eea
where in the second equality we have used cyclic invariance and linearity of the trace operator. 
For incomplete bases $\op P_{N_b}\Liouv[\op\rho]\neq\Liouv[\op\rho]$, and even if 
$\tr\{\Liouv[\op\rho]\}=0$,  $\tr\{\op P_{N_b}\Liouv[\op\rho]\}$ can be non-zero for open systems
as has been shown by Raab {\it et al.} for the Liouvillian in the Lindblad form.\cite{Raab:2000/tca/358}   
In contrast, for isolated systems the population is always conserved because
\bea\nonumber
\tr\{\op P_{N_b}\Liouv[\op\rho]\}&=&-\inumb\tr\{ \mat S^{-1}(\mat H\mat B\mat S -\mat S\mat B\mat H)\} \\
&=& -\inumb\tr\{[\mat H,\mat B]\} =0,
\eea
where $[\mat H]_{kl} = \bra{\varphi_k} \op H \ket{\varphi_l}$. 

For an isolated system, the energy variation with time is given by
\bea\nonumber
\tr\{\dot{\op\rho}\op H\} & = & \tr\{\dt{\mat B}\mat H + \mat B\dot{\mat H}\} \\ \label{eq:Et2}
 & = & \tr\{\dot{\mat H}\mat B - \mat B\mat H\mat S^{-1}\mat\tau - \mat\tau^\dagger\mat S^{-1}\mat H\mat B\},
\eea
where in the second equality we used the $\dot{\mat B}$ definition from \eq{eq:vMCG-DM-B}, 
$[\dot{\mat H}]_{kl} = \bra{\dot{\varphi_k}} \op H \ket{\varphi_l} + \bra{\varphi_k} \op H \ket{\dot{\varphi_l}}$, 
and cyclic invariance of trace. 
Equation (\ref{eq:Et2}) can be further simplified using hermiticity of $\mat B$, $\mat H$, and $\mat S$
\bea \nonumber
 \tr\{\dot{\op\rho}\op H\} & = &  \sum_{\alpha k l} 2\Re\left[{\dot z}_{\alpha k}^*\braOket{\frac{\partial \varphi_k}{\partial  z_{\alpha k}}}{\left[\op 1 - \op P_{N_b}\right]\op H}{\varphi_l}B_{lk}\right] \\ \label{eq:zY}
 &=&-2\Im[\tr\{\dot{\mat z}^\dagger\mat Y\}],
\eea
where
\bea
{[ {\mat Y}]}^\alpha_k=-\inumb\sum_l\braOket{\frac{\partial \varphi_k}{\partial  z_{\alpha k}}}{\left[\op 1 - \op P_{N_b}\right]\op H}{\varphi_l}B_{lk}.
\eea
Using \eq{eq:vMCG-DM-lambda} and the vector notation for $\dot{\mat z}$ and $\mat Y$ we arrive at 
\bea
\tr\{\dot{\op\rho}\op H\} = -2\Im[\tilde{\mat Y}^\dagger\tilde{\mat C}^{-1}\mat Y],\label{eq:not-conserved-E-QME}
\eea
Equation (\ref{eq:vMCG-DM-Y}) simplifies for isolated system to 
\bea
{\tilde Y}^\alpha_k=-\inumb\sum_l\braOket{\frac{\partial \varphi_k}{\partial  z_{\alpha k}}}{\left[\op 1 - \op P_{N_b}\right]\op H}{\varphi_l}[\mat B\mat S\mat B]_{lk}.\hspace{0.2cm}
\eea
According to \eqs{eq:zY} and (\ref{eq:not-conserved-E-QME}), 
energy conservation takes place at least in four special cases:

\begin{enumerate}
\item Time-independent bases, where $\dot{\mat z}=0$.

\item Complete bases, where $\op P_{N_b}=\op 1$, and therefore, $\mat Y=0$.

\item Systems in pure states, where $\mat B\mat S\mat B=\mat B$ (the idempotency condition), 
then $\tilde{\mat Y}=\mat Y$ and using hermiticity of $\tilde{\mat C}$, it is easy to see that 
$\Im[\tilde{\mat Y}^\dagger\tilde{\mat C}^{-1}\mat Y] = \Im[\mat Y^\dagger\tilde{\mat C}^{-1}\mat Y]=0$. 

\item Linear parameterization, $\ket{\varphi_k}=\sum_\alpha \ket{\phi_\alpha} z_{\alpha k}$, 
where $\ket{\phi_\alpha}$ are time-independent basis functions.
In this case, \eq{eq:vMCG-DM-lambda1} becomes 
$\mat C_0\dot{\mat z}\mat B\mat S\mat B=\mat Y_0\mat B\mat S\mat B$,
where $[\mat C_0]_{\alpha\beta}=\langle\phi_\alpha|[\op 1 - \op P_{N_b}]|\phi_\beta\rangle$ 
and $[\mat Y_0]_{\alpha l}=\langle\phi_\alpha|[\op 1 - \op P_{N_b}]\op H|\varphi_l\rangle$, which leads to
$\dot{\mat z}=\mat C_0^{-1}\mat Y_0$, and similarly $\mat Y=\mat Y_0\mat B$.
Equation (\ref{eq:zY}) gives 
$-2\Im[\tr\{\dot{\mat z}^\dagger\mat Y\}]=-2\Im[\tr\{\mat Y_0^\dagger\mat C_0^{-1}\mat Y_0\mat B\}]=0$, 
where we used hermiticity of $\mat C_0$ and $\mat B$, and semi-positivity of $\mat B$.
\end{enumerate}
We would like to conclude this section by emphasizing that a general non-linear time-dependent parameterization 
(e.g., used in \gls{MCTDH} and \gls{vMCG} methods)
of the density matrix will not conserve energy of the isolated system in a mixed state. 

\subsection{Constrained \protect\gls{TDVP} for \protect\gls{NOSSE}}

To resolve the non-conservation problems we will use a problem specific approach: 
The population non-conservation for open system will be corrected by introducing a constraint using 
Lagrange multiplier method.
For the problem of energy non-conservation we reformulate dynamical equations 
using non-stochastic Schr\"odinger equation for a square root of the density. 
This difference in approaches to the two problems stems from an objective to preserve unitarity of the dynamics in the 
isolated system limit. Imposing a constraint ensuring energy conservation breaks the unitarity 
of the isolated system dynamics and may lead to artifacts.~\cite{note:supp}  

To proceed to formulation of \glspl{EOM} for a density square root in open systems, we introduce some notation first.
Matrix $\mat B$ is semi-positive, and therefore, 
there exist a matrix $\mat M$ such that $\mat B=\mat M\mat M^\dagger$. Defining  
\bea\label{eq:ansatz-m}
\ket{m_k} & = & \sum_l \ket{\varphi_l} M_{lk},
\eea 
and using  \eq{eq:ansatz-rho}, we can rewrite the density matrix as
\bea\label{eq:def-m}
\op\rho & = & \sum_{abk}\ket{\varphi_a}M_{ak}M_{bk}^*\bra{\varphi_b} \nonumber\\
& = & \sum_{k}\ket{m_k}\bra{m_k}.
\eea
All states $\{\ket{m_k}\}$ can be arranged in a vector form $\mat m=(\ket{m_1}\ket{m_2}\hdots)$,
such that $\mat m$ represents a density square root because $\op\rho = \mat m\mat m^{\dagger}$. 
In Ref.~\onlinecite{Joubert:2014/jcp/234112} we found \gls{EOM} for $\mat m$ for a system 
with a given Liouvillian operator $\Liouv$ 
\bea\label{eq:NOSSE}
\dot{\mat m} & = & \mat\sqrtLiouv[\mat m],
\eea
where $\mat\sqrtLiouv[\mat m]$ is a vector of states that satisfies a Sylvester equation 
\bea
\mat\sqrtLiouv[\mat m]\mat m^\dagger + \mat m\mat\sqrtLiouv[\mat m]^\dagger = \Liouv[\mat m\mat m^\dagger]. \label{eq:Sylvester}
\eea
Note that for an isolated system $\mat\sqrtLiouv[\mat m]$ reduces to $-iH\mat m$ and \eq{eq:NOSSE} becomes 
an uncoupled set of \gls{TDSE} for each component of $\mat m$. 
This reduction guarantees that application of TDVP to \eq{eq:NOSSE} will be free from energy 
conservation issues arising in the density formalism. 
In order to optimize the parameters of $\mat m$, the Frobenius norm $||\dot{\mat m}-\mat\sqrtLiouv[\mat m]||$ is minimized.

To address non-conservation of the total population for open systems
we introduce a corresponding constraint using the Lagrange multiplier method. The Lagrangian function is 
\bea\label{eq:lag-NOSSE}
\Lag & = & ||\dot{\mat m}-\mat\sqrtLiouv[\mat m]||^2 + \lag\frac{\partial}{\partial t}\tr\{\mat m\mat m^\dagger\},
\eea
where $\lag$ is a Lagrange multiplier.
Minimizing $\Lambda$ with respect to parameters $\mat M=\{M_{kl}\}$, $\mat z$, and $\lag$ leads to a constrained \gls{TDVP}
\bea\label{eq:TDVP-NOSSE}
&\Re\left[\tr\left\{\delta\dot{\mat m}^\dagger\left(\dt{\mat m} - \mat\sqrtLiouv[\mat m] + \lag \mat m\right)\right\}\right] = 0,\\\label{eq:TDVP-NOSSE-cstr}
&\frac{\partial}{\partial t}\tr\{\mat m\mat m^\dagger\} = 0.
\eea
Replacing $\mat m$ by its explicit form [\eq{eq:ansatz-m}] in \eq{eq:TDVP-NOSSE} and \eq{eq:TDVP-NOSSE-cstr} we obtain 
\glspl{EOM} for $\mat M$ and $\mat z$ (details of the derivation are given in App.~\ref{sec:TDVP-NOSSE})
\bea\label{eq:vMCG-NOSSE-M}
\dt{\mat M} & = & \mat S^{-1}\mat K - \mat S^{-1}\mat\tau\mat M\nonumber\\
&& - \frac{1}{2}\tr\{\mat K\mat M^\dagger + \mat M\mat K^\dagger\}\mat M\\\label{eq:vMCG-NOSSE-lambda}
\dt{\mat  z} & = & \mat C^{-1}\mat Y,
\eea
where $K_{kl} = \bra{\varphi_k}\sqrtLiouv_l[\mat m]\rangle$ and
\bea\label{eq:vMCG-NOSSE-C}
 C^{\alpha\beta}_{kl} & = & \braOket{ \frac{\partial\varphi_k}{\partial  z_{\alpha k}} }
                                         { \left[\op 1 - \op P_{N_b} \right] }
                                         { \frac{\partial\varphi_l}{\partial  z_{l\beta}} } [\mat M\mat M^\dagger]_{lk}, \\\label{eq:vMCG-NOSSE-Y}
{{Y}}^\alpha_k & = & \sum_l\bra{ \frac{\partial\varphi_k}{\partial  z_{\alpha k}} }
                  \left[\op 1 - \op P_{N_b}\right]\ket{[\sqrtLiouv_l[\mat m]}M_{kl}^*.
\eea

Using the relation $\dot{\mat B}=\dot{\mat M}\mat M^\dagger+\mat M\dot{\mat M}^\dagger$, \eq{eq:vMCG-NOSSE-M} 
can be recast into an \gls{EOM} for $\mat B$
\bea\label{eq:vMCG-NOSSE-B}
\dt{\mat B} & = & \mat S^{-1}\mat L\mat S^{-1} - \left(\mat S^{-1}\mat\tau\mat B + \mat B\mat\tau^\dagger\mat S^{-1}\right) - \tr\{\mat S^{-1}\mat L\}\mat B, \nonumber\\
\eea
where we employed the Sylvester equation [\eq{eq:Sylvester}] in the matrix form, $\mat L=\mat K\mat M^\dagger\mat S + \mat S\mat M\mat K^\dagger$.

Now, it is easy to show that Eqs.~(\ref{eq:vMCG-NOSSE-lambda}-\ref{eq:vMCG-NOSSE-B}) conserve all required quantities.
Taking the trace of $\dt{\op\rho}=\frac{\partial}{\partial t}\sum_{kl}\ket{\varphi_k}B_{kl}\bra{\varphi_l}$ and using \eq{eq:vMCG-NOSSE-B} gives
\bea\label{eq:app-rho-conserved-NOSSE-os}
\frac{\partial}{\partial t}\tr\{\op\rho\} & = & \tr\{\dt{\mat B}\mat S+(\mat\tau+\mat\tau^\dagger)\mat B\} \nonumber\\
 & = & \tr\big\{\mat S^{-1}\mat L-\tr\{\mat S^{-1}\mat L\}\mat B\mat S\big\}=0,
\eea
where the last equality holds since $\tr\{\op\rho\}=\tr\{\mat B\mat S\}=1$.
Therefore, the trace is conserved for any Liouvillian $\Liouv[\op\rho]$.
If the system is isolated, $\Liouv[\op\rho]=-\inumb[\op H, \op\rho]$ in \eq{eq:vMCG-NOSSE-B}, 
it is possible to show that a trace of any power of $\op\rho$
is also conserved
\bea
\frac{\partial}{\partial t}\tr\{\op\rho^n\} & = & \tr\{\dt{\mat B}\mat S(\mat B\mat S)^{n-1} + \mat B(\mat\tau+\mat\tau^\dagger)(\mat B\mat S)^{n-1}\} \nonumber\\
&&  \hspace{-2cm}=-\inumb\tr\big\{\mat S^{-1}(\mat H\mat B\mat S-\mat S\mat B\mat H)(\mat B\mat S)^{n-1} \nonumber\\
&&  \hspace{-0.75cm}-\tr\{\mat S^{-1}(\mat H\mat B\mat S-\mat S\mat B\mat H)\}(\mat B\mat S)^n\big\}=0.\label{eq:app-rhon-conserved-NOSSE-cs}
\eea
This illustrates unitary character of dynamics. 
The energy is also conserved for the isolated system limit
\bea
\frac{\partial}{\partial t}\tr\{\op\rho\op H\} & = & \tr\{\dt{\mat B}\mat H + \mat B\dot{\mat H}\} \nonumber\\
&& \hspace{-1.5cm}=\tr\{\mat B\dot{\mat H} - \mat S^{-1}\mat\tau\mat B\mat H - \mat H\mat B\mat\tau^\dagger\mat S^{-1}\} \nonumber\\
&& \hspace{-1.5cm}=-2\Im[\dot{\mat z}^\dagger\mat Y] = -2\Im[\mat Y^\dagger\mat C^{-1}\mat Y] = 0,\label{eq:app-rhoH-conserved-NOSSE-cs}
\eea
where we invoke the hermiticity of $\mat C$ to see that $\mat Y^\dagger\mat C^{-1}\mat Y$ does not have imaginary component. 

\section{Numerical examples}
\label{sec:results}

\subsection{Model}

To illustrate performance of our developments for the case where 
system quantum effects are affected by interaction with environment, we will consider the simplest 
\gls{LVC} model\cite{Koppel:1984/acp/59} of crossing surfaces.~\cite{Migani:2004/271,Domcke:2012/arpc/325}
The system contains two electronic states, donor $\ket{D}$ and acceptor $\ket{A}$, which are coupled through two nuclear \glspl{DOF}
\bea\label{eq:2D-Ham}
\op H & = & \sum_{\alpha=1}^2\frac{\omega_\alpha}{2}\left(\op p_\alpha^2 + \op x_\alpha^2\right)\Big[\ket{D}\bra{D}+\ket{A}\bra{A}\Big]\\
&&- d\op x_1\Big[\ket{D}\bra{D}-\ket{A}\bra{A}\Big] + c\op x_2\Big[\ket{D}\bra{A}+\ket{A}\bra{D}\Big],\nonumber
\eea
where $\omega_\alpha$ is the frequency of the coordinate $x_\alpha$, and $d$ and $c$ are diabatic coupling constants.
Numerical values of the parameters are taken from Ref.~\onlinecite{Ryabinkin:2014/jcp/214116} for the two-dimensional model of the bis(methylene) adamantyl radical cation: $\omega_1=7.743\cdot10^{-3}$ a.u., $\omega_2= 6.680\cdot10^{-3}$ a.u., $d=5.289\cdot10^{-3}$ a.u., $c=9.901\cdot10^{-4}$ a.u.
The dissipative part is introduced through bilinear coupling of the system coordinates $x_\alpha$ with coordinates of 
harmonic bath oscillators. System-bath couplings are taken small enough so that the second order perturbation theory 
can be applied for the system-bath interaction in conjunction with the rotating wave and Markovian approximations. The resulting Liouvillian~\cite{Takagahara:1978/jpsj/728,Wolfseder:1995/cpl/370} has a Lindblad form~\cite{Lindblad:1976/cmp/119}
\bea\nonumber
\Liouv[\op\rho] & = & -\inumb[\op H,\op\rho] + h_1\sum_{\alpha=1}^2\left(2\op Z_\alpha\op\rho{\op Z_\alpha}^\dagger-{\op Z_\alpha}^\dagger\op Z_\alpha\op\rho-\op\rho{\op Z_\alpha}^\dagger\op Z_\alpha\right)\\
       &   & + h_2\sum_{\alpha=1}^2\left(2{\op Z_\alpha}^\dagger\op\rho\op Z_\alpha-\op Z_\alpha{\op Z_\alpha}^\dagger\op\rho-\op\rho\op Z_\alpha{\op Z_\alpha}^\dagger\right),\label{eq:DM-Lindblad}
\eea
where $h_1$ and $h_2$ are parameters that depend on the bath temperature, and $\op Z_\alpha$ 
operators are defined as 
\bea
\op Z_1 & = & \left(\op a_1-\frac{d}{\omega_1\sqrt{2}}\right)\ket{D}\bra{D} + \left(\op a_1+\frac{d}{\omega_1\sqrt{2}}\right)\ket{A}\bra{A} \nonumber\\
\op Z_2 & = & \op a_2\Big(\ket{D}\bra{D} + \ket{A}\bra{A}\Big),\label{eq:Z}
\eea
with annihilation operators $\op a_\alpha=(\op x_\alpha+\inumb\op p_\alpha)/\sqrt{2}$.
For $\Liouv[\op\rho]$ in the Lindblad form, $\mat\sqrtLiouv[\mat m]$ is derived in Ref.~\onlinecite{Joubert:2014/jcp/234112}, 
and for our model is given by
\bea\nonumber
\mat\sqrtLiouv[\mat m] & = & -\inumb\op H\mat m + h_1\sum_{\alpha=1}^2\left(\op Z_\alpha\op\rho{\op Z_\alpha}^\dagger(\mat m^\dagger)^{-1}-{\op Z_\alpha}^\dagger\op Z_\alpha\mat m\right)\\
       &   & + h_2\sum_{\alpha=1}^2\left({\op Z_\alpha}^\dagger\op\rho\op Z_\alpha(\mat m^\dagger)^{-1}-\op Z_\alpha{\op Z_\alpha}^\dagger\mat m\right).\label{eq:NOSSE-Lindblad}
\eea

\subsection{Numerical details}

We apply our constrained \gls{TDVP} to the \glsfirst{vMCG} ansatz.~\cite{Worth:2004/fd/307}
In \gls{vMCG}, the time-dependent basis consist of products 
\bea
\ket{\varphi_k} & = & \ket{\sigma_k}\ket{g_k} \label{eq:basis-prod}
\eea
of discrete electronic parts
$\ket{\sigma_k}= \ket{D},\ket{A}$ and spatial 
two-dimensional moving Gaussian parts $\ket{g_k}$ parametrized as follows
\bea\label{eq:gaussians}
\braket{\mat x}{g_k} & = & \exp\left(-\mat x^T\mat A_k\mat x + \mat\xi_k^T\mat x\right),
\eea
where $\mat x$ is a vector of the mass- and frequency-weighted nuclear coordinates,
$\mat A_k$ are frozen widths ($\dot{\mat A}_k=0$) and $\mat\xi_k$ are parameters encoding the 
position (${\rm Re}[\mat \xi_k]$) and momentum (${\rm Im}[\mat \xi_k]$) of a Gaussian.
This choice of parameterization satisfy the analyticity condition $\partial\ket{\varphi_k}/\partial \xi_{k\alpha}^*=0$.

For reasons of numerical stability it is more convenient to 
work with \glspl{CS} of harmonic oscillators rather than frozen Gaussians.\cite{Izmaylov:2013/jcp/104115,Saita:2012di} 
However, \glspl{EOM} derived from the McLachlan TDVP expressions for \glspl{CS}
can have energy conservation issues because \glspl{CS} do not satisfy the analyticity condition. 
On the other hand, \gls{TDVP} \glspl{EOM} for frozen Gaussians  
can be projected onto the basis of \glspl{CS}, $\{\ket{\mat z_k}\}$, of harmonic oscillators 
with frequencies $\omega_\alpha$ by multiplying \eq{eq:gaussians} with
$V_{kk}=e^{-\mat\xi_k^T\Re[\mat\xi_k]/2}/\pi$ and setting $\mat A_k={\bf 1}_{2}/2$. 
The transformation of Gaussians $\ket{\mat z_k}= V_{kk}\ket{g_k}$
leads to a transformation of the matrix $\mat B=\mat V\tilde{\mat B}\mat V^\dagger$, 
where $[\mat V]_{kl}=V_{kk}\delta_{kl}$. 
In the \gls{CS} representation variational parameters become $z_{\alpha k}=\xi_{k\alpha}/\sqrt{2}$, 
they correspond to eigenvalues of the annihilation operator $\op a_\alpha$ for \glspl{CS} $\ket{\mat z_k}$ 
\bea\label{eq:cs}
  \op a_\alpha\ket{\mat z_k} & = & z_{\alpha k}\ket{\mat z_k}.
\eea
Upon the $\mat V$ transformation, the structure of the Eqs.~(\ref{eq:vMCG-NOSSE-lambda}-\ref{eq:vMCG-NOSSE-B}) 
remains the same, thus, we will use these equations to propagate parameters of \glspl{CS}.

The population loss for an open system is given by \eq{eq:noTd}. 
In the particular case of the Liouvillian defined by \eq{eq:DM-Lindblad} in the basis of \glspl{CS}, 
the expression for the population loss becomes
\bea\label{eq:trdr}
\tr\{\dot{\op\rho}\} & = & 2\sum_{\alpha=1}^2\tr\{ [h_1 \op Z_\alpha^\dagger(\op P_{N_b} - 1)\op Z_\alpha \\[-0.35cm]\nonumber
 & &\hspace{2.25cm} + h_2 \op Z_\alpha(\op P_{N_b} - 1){\op Z_\alpha}^\dagger]\op\rho\}.
 \eea
 Equation~(\ref{eq:trdr}) 
 can be further simplified using the definitions of $\op Z_\alpha$ [\eq{eq:Z}], $\op\rho$ [\eq{eq:ansatz-rho}], $\op P_{N_b}$, 
 and applying \eq{eq:cs} 
\bea
 \tr\{\dot{\op\rho}\} & = & 2 h_2\sum_{\alpha=1}^2\tr_N\{\op a_\alpha(\op P_{N_b} - 1){\op a_\alpha}^\dagger \nonumber\\[-0.35cm]
 & &\hspace{2cm} \times [\braOket{D}{\op\rho}{D}+\braOket{A}{\op\rho}{A}]\},\label{eq:rho-not-conserved-Lindblad}
\eea
where $\tr_N\{\hdots\}$ is the trace over system nuclear \glspl{DOF}.
Thus, the $h_1$ containing term disappears and the population loss takes place only if $h_2\neq0$.

The initial system \gls{DM} $\op\rho$ is taken as a Boltzmann distribution $\op\rho_{\rm B}$ in the donor electronic state
\bea\label{eq:boltz}
\op\rho_{\rm B}=\frac{e^{-\op P_D\op H\op P_D/k_BT}\op P_D}{\tr\{e^{-\op P_D\op H\op P_D/k_BT}\op P_D\}},
\eea
where $\op P_D=\ket{D}\bra{D}$ is the projector on the donor electronic state, 
$k_B$ is the Boltzmann constant and $T$ is the temperature.
An approximate initial density matrix $\op\rho(0)$ is obtained by minimizing 
\bea
\tilde{\Lambda} = ||\mat m-\mat m_{\rm B}||^2+\lambda[\tr\{\mat m\mat m^{\dagger}\}-1],
\eea
where $\mat m_{\rm B}$ is a square root of $\op\rho_{\rm B}$. 
This minimization is equivalent to a two-step procedure: first, maximizing $\tr\{\op\rho_{\rm B}\op P_{N_b}\}$ with 
respect to the parameters of the basis, and second, constructing the initial density 
in the optimized basis as $\op\rho(0)=\op P_{N_b}\op\rho_{\rm B}\op P_{N_b}/\tr\{\op\rho_{\rm B}\op P_{N_b}\}$.

Solving the \glspl{EOM}~(\ref{eq:vMCG-NOSSE-lambda}-\ref{eq:vMCG-NOSSE-B}) 
requires inversion of the $\mat S$ and $\mat C$ matrices, 
these matrices can become singular and lead to numerical difficulties.~\cite{Burghardt:2003/jcp/5364} 
This problem is addressed by applying a regularization procedure~\cite{Meyer:2009/book} on 
singular eigenvalues ($s$) before the matrix inversion: $s \rightarrow s+\varepsilon e^{-s/\varepsilon}$,
where $\varepsilon=10^{-6}$ is the chosen threshold. 
This procedure provides a faster and more stable propagation with a numerical precision of the order of $\varepsilon$.
The limiting step of our implementation is the inversion of the $\mat C$ matrix. The computational 
cost of this inversion scales cubically with the dimensionality of $\mat C$. 
For the system with $d$ \gls{DOF} described by $N_b$ Gaussians, the dimensionality of $\mat C$
is $2 N_b d$ (where the factor of two accounts for position and momentum parameters of Gaussians),
and thus the computational cost scales as $8 N_b^3 d^3$.

Simulations with time-dependent basis are compared with exact simulations that are done by projecting the \gls{QME} 
onto the direct-product basis of the two-dimensional harmonic oscillator. 
360 static basis functions are employed to obtain converged results that we will denote as $\op\rho_e(t)$.

\subsection{Isolated system simulations}
\label{sec:res-close}

\begin{figure}
  \centering
  \includegraphics[width=0.5\textwidth]{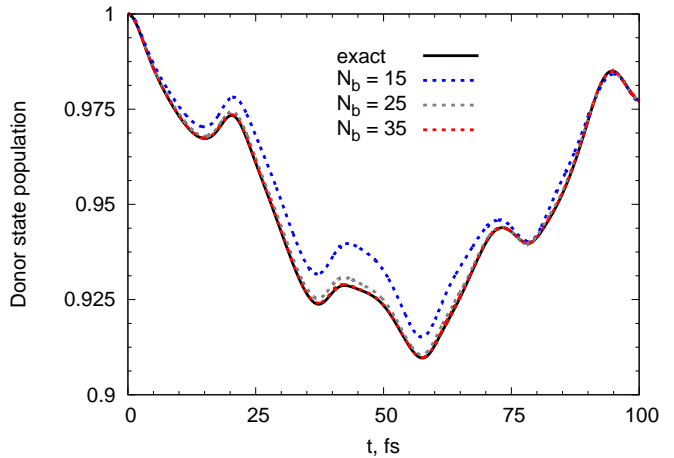}
  \caption{Isolated system: Population dynamics of the donor electronic state $\tr\{\op\rho(t)\op P_D\}$ for the exact and \protect\gls{vMCG} dynamics based on constrained \protect\gls{NOSSE} [Eqs.~\eqref{eq:vMCG-NOSSE-C} and~\eqref{eq:vMCG-NOSSE-Y}] formalism with different number $N_b$ of \protect\glspl{CS}.}
  \label{fig:pop-close}
\end{figure}

For an isolated system where $h_1=h_2=0$, $\Liouv[\op\rho]=-\inumb[\op H,\op\rho]$, and $\mat\sqrtLiouv[\mat m]=-\inumb\op H\mat m$,
we choose the temperature of the initial Boltzmann distribution to be $T=1000$ K (or $T=0.47\omega_1=0.41\omega_2$).
Figure~\ref{fig:pop-close} shows population dynamics of the donor state, $\tr\{\op\rho\op P_D\}$, simulated with the \gls{NOSSE}-based formalism.
A good agreement with the exact propagation is obtained already for $25$ \glspl{CS}, 
and results of our method converges to those of the exact propagation with $35$ \glspl{CS}.
\Glspl{RMSD} from the exact dynamics for various properties are defined as
\bea
\label{eq:RMSD}
{\rm RMSD}(A) & = & \sqrt{\frac{1}{t_f}\int_0^{t_f}dt\tr\{\op A[\op\rho(t)-\op\rho_e(t)]\}^2},
\eea
where $t_f=100$ fs, and $\op A=\op 1,~\op P_D$, and $\op H$ for the total population ($P_T$), 
the donor state population ($P_D$), and the system energy ($E$), respectively. 
Table~\ref{tab:close} shows that the system energy in 
\gls{QME}-based simulations is not well conserved. 
As a result, the corresponding donor population dynamics have larger deviations than those in the NOSSE formalism.
\begin{table}[!b]
  \centering
  \caption{\protect\glspl{RMSD} of the donor population and energy  
  with the \protect\gls{NOSSE} and \protect\gls{QME} approaches for different number $N_b$ of CSs.}
  \begin{ruledtabular}
    \begin{tabular}{@{}l@{}c@{}c@{}c@{}}
$N_b$      &       $15$        &       $25$        &       $35$       \\
\hline
\multicolumn{4}{c}{ RMSD($E$), in units of $\omega_1$} \\
NOSSE      & $<10^{-5}$ & $<10^{-5}$ & $<10^{-5}$\\
QME  &  $2.8\cdot10^{-3}$  &  $2.3\cdot10^{-3}$  &  $2.4\cdot10^{-3}$\\
 \multicolumn{4}{c}{ RMSD($P_D$)} \\
NOSSE      & $5.9\cdot10^{-3}$ & $1.1\cdot10^{-3}$ & $1.7\cdot10^{-4}$\\
 QME       & $6.1\cdot10^{-3}$ & $1.1\cdot10^{-3}$ & $4.9\cdot10^{-4}$
    \end{tabular}
  \end{ruledtabular}
  \label{tab:close}
\end{table}

The coupling between two electronic states is relatively weak and the population dynamics can be analyzed 
using perturbation theory. Fast scale oscillations are produced from transitions between $(m,n)_D$ and $(m,n\pm1)_A$
levels due to the linear coupling [$c \op x_2$ in \eq{eq:2D-Ham}], here, the first and the second numbers correspond to the numbers of 
vibrational quanta along the tuning ($x_1$) and coupling ($x_2$) modes, and the subscripts denote the electronic states. 
Vibrational levels $(m,n)_D$ and $(m,n\pm1)_A$ 
are separated by the energy difference $\pm\omega_2$, which translates into the population oscillation period 
$t_p = 2\pi/\omega_2 = 23$ fs. There are also slower time-scale oscillations that appears as a result of transitions 
between the $(m,n)_D$ and $(m\mp1,n\pm1)_A$ levels. The tuning mode does not contribute to the coupling 
between electronic states, but it shifts minima of these states so that levels' couplings are modulated by 
the Franck-Condon factors. The energy difference between the $(m,n)_D$ and $(m\mp1,n\pm1)_A$ levels 
is $\pm(\omega_1-\omega_2)$ that results in $143$ fs time difference between two subsequent maxima of slow population oscillations.  

\subsection{Open system simulations}
\label{sec:res-open}

\begin{figure}
  \centering
  \includegraphics[width=0.5\textwidth]{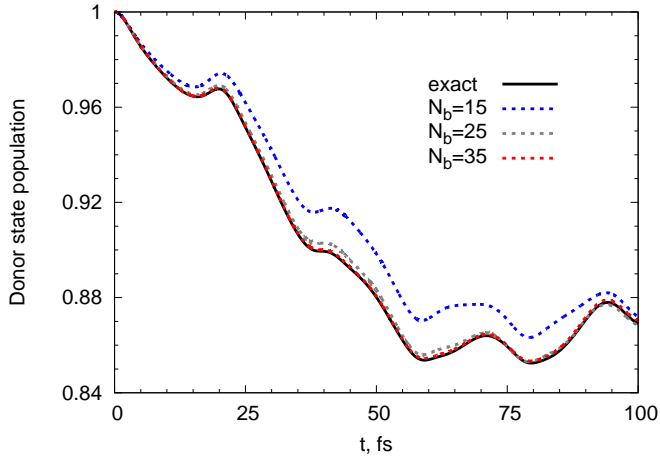}
  \caption{Open system: Population dynamics of the donor electronic state $\tr\{\op\rho(t)\op P_D\}$ for the exact and \protect\gls{vMCG} dynamics based on constrained \protect\gls{NOSSE} [Eqs.~\eqref{eq:vMCG-NOSSE-C} and~\eqref{eq:vMCG-NOSSE-Y}] formalism with different number $N_b$ of \protect\glspl{CS}.}
  \label{fig:pop-warm}
\end{figure}

We simulate the system ($T=1000$ K) located in the donor well and interacting with a harmonic bath according to 
\eq{eq:DM-Lindblad} with $h_1=0$ and $h_2=3.675\cdot10^{-4}$ a.u. 
For $h_2\neq0$ the trace of $\op\rho$ varies if no constraints are applied [see \eq{eq:rho-not-conserved-Lindblad}].
In the constrained \gls{NOSSE}-based \gls{vMCG} simulations, the system density trace is conserved up to $2\cdot10^{-5}$.
The population dynamics of the donor state $\tr\{\op\rho(t)\op P_D\}$ is given in Fig.~\ref{fig:pop-warm}, it illustrates 
that the constrained \gls{NOSSE} \gls{vMCG} formalism gives the exact dynamics for $35$ \glspl{CS}.
A comparison of deviations from the exact dynamics in 
trace-constrained and unconstrained \gls{NOSSE} simulations 
is given in Table~\ref{tab:open}.  As expected, the unconstrained \gls{NOSSE} donor populations 
have larger deviations than those of the constrained formalism.
 Although generally there is no notable computational cost difference between constrained and 
unconstrained schemes, in some cases the constrained scheme has better efficiency due to 
more regularly behaving solutions of the corresponding EOM. Thus, all open system calculations
similar to the ones reported here should be done with the constrained formalism.   
\begin{table}[!b]
  \centering
  \caption{\protect\glspl{RMSD} of the total and donor populations 
  with the constrained and unconstrained \protect\gls{NOSSE} approaches for different number $N_b$ of CSs.}
  \begin{ruledtabular}
    \begin{tabular}{@{}l@{}c@{}c@{}c@{}}
   $N_b$      &       $15$        &       $25$        &       $35$       \\
   \hline
 \multicolumn{4}{c}{ RMSD($P_T$)} \\ 
  Constrained & $<10^{-5}$ & $<10^{-5}$ & $<10^{-5}$\\
Unconstrained & $8.8\cdot10^{-2}$ & $2.2\cdot10^{-2}$ & $6.5\cdot10^{-3}$ \\
 \multicolumn{4}{c}{ RMSD($P_D$)} \\
  Constrained & $1.3\cdot10^{-2}$ & $1.9\cdot10^{-3}$ & $6.2\cdot10^{-4}$\\
Unconstrained & $6.8\cdot10^{-2}$ & $1.8\cdot10^{-2}$ & $5.0\cdot10^{-3}$
    \end{tabular}
  \end{ruledtabular}
  \label{tab:open}
\end{table}

The fast-scale oscillations ($23$ fs period) observed in the isolated system case are also present in Fig.~\ref{fig:pop-warm}, 
and their nature is exactly the same. However, due to influence of the environment, 
the long term dynamics has changed in the open system case. The origin of the irreversible decay 
of the donor population is a two-step-resonance process: $(0,0)_D \rightarrow (0,1)_D \rightarrow (0,0)_A$,
where the first step is driven by the environment excitation of the system within the donor state, 
while the second step is due to the linear coupling between the electronic states. 
Since there is no energy difference between the initial $(0,0)_D$ and the 
final $(0,0)_A$ states, this process  leads to the larger population transfer observed in Fig.~\ref{fig:pop-warm}. 

\section{Concluding remarks}
\label{sec:conclusion}
In this work we exposed the energy and total population non-conservation problems that 
occur in the density matrix \gls{TDVP} formalism.  The proposed constrained \gls{TDVP}-\gls{NOSSE} 
approach resolves the non-conservation issues by combining the Lagrange multiplier method 
to preserve the total population and the \gls{NOSSE} formalism to reproduce unitary dynamics in 
the isolated system limit. 
As illustrated on a model surface-crossing problem, the developed approach can significantly 
reduce the number of propagated parameters for simulating non-adiabatic dynamics while conserving
all required physical quantities. Moreover, even larger efficiency improvements compare 
to standard techniques are expected in molecular problems with a large number of nuclear \gls{DOF}. 
Although the current implementation of the constrained \gls{TDVP}-\gls{NOSSE} approach 
has been illustrated using \gls{vMCG} ansatz, working equations can be easily used to derive other system density 
parametrizations (e.g., \gls{MCTDH}). For a better numerical stability we have 
used a basis of \glspl{CS} that are localized in space.
Thus our implementation is perfectly suited for a combination
with the on-the-fly evaluation of electronic potential energy surfaces.
\cite{Bennun:2000/jpca/5161, Saita:2012di,Worth:2008/mp/2077}
From the application prospective, our developments open a new venue of the on-the-fly 
photochemistry with incoherent light. The sun light is incoherent, hence, realistic modelling of chemical 
photostability, solar energy harvesting and utilization requires introducing light as the
system environment in a mixed state, which can be easily handled by the current approach.

\section*{Acknowledgements}

L.J.D. thanks Ilya Ryabinkin for useful discussions and 
the European Union Seventh Framework Programme (FP7/2007-2013) for 
financial support under grant agreement PIOF-GA-2012-332233.
A.F.I. acknowledges funding from the Natural Sciences and Engineering Research Council 
of Canada (NSERC) through the Discovery Grants Program.

\section{Appendix: derivation of the equations of motion for the \protect\gls{TDVP} applied to \protect\gls{QME}}
\label{sec:TDVP-QME}

Main steps in derivation of Eqs.~(\ref{eq:vMCG-DM-B})~-~(\ref{eq:vMCG-DM-lambda}) are detailed below.
Considering the parameterization of the \gls{DM} in \eq{eq:ansatz-rho},
variation of $\delta\op\rho$ in \eq{eq:TDVP-QME} 
can be split into contributions from $B_{kl}$, $ z_{\alpha k}$, and $z_{\alpha k}^*$
\bea\label{eq:rho_variations-app}
\delta\op\rho & = & \sum_{kl}\frac{\partial\op\rho}{\partial B_{kl}}\delta B_{kl} + \sum_{k\alpha}\Big[\frac{\partial\op\rho}{\partial  z_{\alpha k}}\delta z_{\alpha k} + \frac{\partial\op\rho}{\partial  z_{\alpha k}^*}\delta z_{\alpha k}^*\Big].\nonumber\\
\eea
Owing to independent character of $\delta B_{kl}$, $\delta z_{\alpha k}$, and $\delta z^*_{k\alpha}$ variations in 
\eq{eq:rho_variations-app}, \eq{eq:TDVP-QME} represents a system of equations
\bea\label{eq:vMCG-QME-Bkl-app}
&\tr\left\{\frac{\partial\op\rho}{\partial B_{kl}}\left(\dt{\op\rho} - \Liouv[\op\rho]\right)\right\} = 0,\\\label{eq:vMCG-QME-lambdakalpha-app}
&\tr\left\{\frac{\partial\op\rho}{\partial z_{\alpha k}}\left(\dt{\op\rho} - \Liouv[\op\rho]\right)\right\} = 0,\\\label{eq:vMCG-QME-lambdakalpha-2-app}
&\tr\left\{\frac{\partial\op\rho}{\partial z_{\alpha k}^*}\left(\dt{\op\rho} - \Liouv[\op\rho]\right)\right\} = 0.
\eea
Replacing $\op\rho$ by its explicit form [\eq{eq:ansatz-rho}] into \eq{eq:vMCG-QME-Bkl-app} leads to
\bea\nonumber
&&\sum_{ab}\frac{\partial B_{ba}}{\partial B_{kl}}\bra{\varphi_b}\Big\{\sum_{cd}\Big(\ket{\varphi_c}\dt{B}_{cd}\bra{\varphi_d} +\ket{\dt{\varphi}_c}B_{cd}\bra{\varphi_d}\\[-0.3cm]\nonumber
&& \hspace{3cm} +\ket{\varphi_c}B_{cd}\bra{\dt{\varphi}_d}\Big) - \Liouv[\op\rho]\Big\}\ket{\varphi_a}\\
& = & \sum_{cd}\Big(S_{kc}\dt{B}_{cd}S_{dl} +\tau_{kc}B_{cd}S_{dl} +S_{kc}B_{cd}\tau_{ld}^* \Big) -L_{kl} = 0,\nonumber\\\label{eq:elem-B}
\eea
where $S_{kc}=\braket{\varphi_k}{\varphi_c}$ is the overlap matrix between 
time-dependent basis functions, $\tau_{kc}=\braket{\varphi_k}{\dot{\varphi}_c}$, and $L_{kl}=\braOket{\varphi_k}{\Liouv[\op\rho]}{\varphi_l}$.
Using the matrix notation, \eq{eq:elem-B} is completely equivalent to \eq{eq:vMCG-DM-B}.

Equations (\ref{eq:vMCG-QME-lambdakalpha-app}) and (\ref{eq:vMCG-QME-lambdakalpha-2-app}) are complex 
conjugates of each other, and therefore, constitute one unique stationary condition.
Replacing $\op\rho$ by its explicit form [\eq{eq:ansatz-rho}] into \eq{eq:vMCG-QME-lambdakalpha-app} leads to
\bea\nonumber
&&\sum_{ab}B_{ab}\bra{\frac{\partial \varphi_b}{\partial z_{\alpha k}}}\Big\{\sum_{cd}\Big(\ket{\dt{\varphi}_c}B_{cd}\bra{\varphi_d} +\ket{\varphi_c}B_{cd}\bra{\dt{\varphi}_d}\\[-0.3cm]\nonumber
&& \hspace{3.5cm} +\ket{\varphi_c}\dt{B}_{cd}\bra{\varphi_d}\Big) - \Liouv[\op\rho]\Big\}\ket{\varphi_a}=0.
\eea
Substituting ${\dot B}_{cd}$ by \eq{eq:vMCG-DM-B} we obtain
\bea\nonumber
&&\sum_{a}\bra{\frac{\partial \varphi_k}{\partial  z_{\alpha k}}}\Big\{\sum_{cd}\Big(\ket{\dt{\varphi}_c}B_{cd}S_{da} +\ket{\varphi_c}B_{cd}\braket{\dt{\varphi}_d}{\varphi_a}\\[-0.2cm]\nonumber
&& \hspace{1cm} +\ket{\varphi_c}[\mat S^{-1}\mat L\mat S^{-1} - \mat S^{-1}\mat\tau\mat B - \mat B\mat\tau^\dagger\mat S^{-1}]_{cd}S_{da}\Big)\\[-0.2cm]\nonumber
&& \hspace{5cm} -\Liouv[\op\rho]\ket{\varphi_a}\Big\}B_{ak}\\\nonumber
&&=\sum_{a}\bra{\frac{\partial \varphi_k}{\partial  z_{\alpha k}}}\Big\{\sum_{cd}\big(\op 1 - \op P_{N_b}\big)\ket{\dt{\varphi}_c} B_{cd}S_{da}\Big)\\[-0.3cm]\nonumber
&& \hspace{4cm} + \big(\op P_{N_b} - \op 1\big) \Liouv[\op\rho]\ket{\varphi_a}\Big\}B_{ak}\\ \label{eq:tder}
&&=0,
\eea
where we used the definition of the projector $\op P_{N_b}=\sum_{kl}\ket{\varphi_k}[\mat S^{-1}]_{kl}\bra{\varphi_l}$.
The time derivative in \eq{eq:tder} can be expanded using the chain rule 
in terms of the basis set parameters $\partial/\partial t=\sum_{l\beta}\dot{ z}_{l\beta}\partial/\partial z_{l\beta}$, thus
\bea\nonumber
&&\sum_{l\beta}\bra{\frac{\partial \varphi_k}{\partial z_{\alpha k}}}\Big[\op 1 - \op P_{N_b}\Big]\ket{\frac{\partial \varphi_l}{\partial z_{l\beta}}} [\mat B\mat S\mat B]_{lk}\dot{ z}_{l\beta}\\\nonumber
&&\hspace{1.5cm}-\bra{\frac{\partial \varphi_k}{\partial z_{\alpha k}}}\Big[\op 1 - \op P_{N_b}\Big]\Liouv[\op\rho]\ket{\varphi_l}[\mat B]_{lk}\\
&&=\sum_{l\beta}\tilde C_{kl}^{\alpha\beta}\dot{ z}_{l\beta} - \tilde Y_k^\alpha =0, \label{eq:elem-z}
\eea
that is \eq{eq:vMCG-DM-lambda1} with definitions given in Eqs.~(\ref{eq:vMCG-DM-C}) and (\ref{eq:vMCG-DM-Y}).

\section{Appendix: derivation of the equations of motion for the constrained \protect\gls{TDVP} applied to \protect\gls{NOSSE}}
\label{sec:TDVP-NOSSE}

Here we provide main steps in derivation of Eqs.~(\ref{eq:vMCG-NOSSE-M})~-~{\ref{eq:vMCG-NOSSE-Y}).
Starting from the constrained \gls{TDVP} \eq{eq:TDVP-NOSSE} we note that variations
\bea
\delta\dot{\mat m}^\dagger & = & \sum_{kl} \frac{\partial\mat m^\dagger}{\partial M_{kl}^*}\delta \dot{M}_{kl}^* + 
\sum_{k\alpha} \frac{\partial\mat m^\dagger}{\partial  z_{\alpha k}^*}\delta \dot{z}_{\alpha k}^*
\eea
and 
\bea\label{eq:vm}
\delta\mat m^\dagger & = & \sum_{kl} \frac{\partial\mat m^\dagger}{\partial M_{kl}^*}\delta M_{kl}^* + 
\sum_{k\alpha} \frac{\partial\mat m^\dagger}{\partial  z_{\alpha k}^*}\delta z_{\alpha k}^*
\eea
span the same tangent space of $\mat m(t)$ because variations of parameters ($\delta M_{kl}^*$, $\delta z_{\alpha k}^*$) 
and their time derivatives ($\delta \dot{M}_{kl}^*$, $\delta \dot{z}_{\alpha k}^*$) are completely arbitrary.
Therefore, \eq{eq:TDVP-NOSSE} can be substituted as 
\bea\label{eq:Re}
\Re\left[\tr\left\{\delta\mat m^\dagger\left(\dt{\mat m} - \mat\sqrtLiouv[\mat m] + \lag \mat m\right)\right\}\right] = 0.
\eea
The parameterization of $\mat m(t)$ is chosen so that the basis functions $\ket{\varphi_k}$ are analytic with respect to their parameters, 
$\partial\ket{\varphi_k}/\partial z_{\alpha k}^*=0$. This analyticity allows us to extend the zero condition in \eq{eq:Re} from a real part 
to the total trace expression\cite{Broeckhove:1988/cpl/547} 
\bea\label{eq:TDVP-DF}
\tr\left\{\delta\mat m^\dagger\left(\dt{\mat m} - \mat\sqrtLiouv[\mat m] + \lag \mat m\right)\right\} = 0.
\eea
The variation $\delta\mat m^\dagger$ is split into contribution from $M_{kl}$ and $ z_{\alpha k}$ [\eq{eq:vm}], and 
using independence and arbitrariness of variations for different parameters, \eq{eq:TDVP-DF} is rewritten as a set of equations
\bea\label{eq:vMCG-NOSSE-Mkl}
\tr\left\{\frac{\partial\mat m^\dagger}{\partial M_{kl}^*}\left(\dt{\mat m} - \mat\sqrtLiouv[\mat m] + \lag \mat m\right)\right\} = 0,\\\label{eq:vMCG-NOSSE-lambdakalpha}
\tr\left\{\frac{\partial\mat m^\dagger}{\partial z_{\alpha k}^*}\left(\dt{\mat m} - \mat\sqrtLiouv[\mat m] + \lag \mat m\right)\right\} = 0.
\eea

Replacing $\mat m^\dagger$ by its explicit form [\eq{eq:ansatz-m}] in \eq{eq:vMCG-NOSSE-Mkl} leads to
\bea\nonumber
&&\sum_{ab}\frac{\partial M_{ba}^*}{\partial M_{kl}^*}\bra{\varphi_b}\Big\{\sum_c\Big(\ket{\varphi_c}\dt{M}_{ca} +\ket{\dt{\varphi}_c}M_{ca}\\[-0.3cm]\nonumber
&& \hspace{4cm} + \lag \ket{\varphi_c}M_{ca}\Big) - \ket{\sqrtLiouv_a[\mat m]}\Big\}\\ \label{eq:M}
& = & \sum_c\Big(S_{kc}\dt{M}_{cl} +\tau_{kc}M_{cl} + \lag S_{kc}M_{cl}\Big)- K_{kl} = 0,
\eea
or using a matrix notation for \eq{eq:M} gives
\bea\label{eq:Mdot}
\dt{\mat M} & = & \mat S^{-1}\mat K - \mat S^{-1}\mat\tau\mat M - \lag \mat M.
\eea
Inserting \eq{eq:Mdot} into the constraint \eq{eq:TDVP-NOSSE-cstr} gives the expression of the Lagrange multiplier
\bea\label{eq:lag}
\lag & = & \frac{\tr\{\mat M\mat K^\dagger+\mat K\mat M^\dagger\}}{2\tr\{\mat M^\dagger\mat S\mat M\}}\nonumber\\
& = & \frac{1}{2}\tr\{\mat M\mat K^\dagger+\mat K\mat M^\dagger\},
\eea
where the second equality arises because the $\tr\{\mat M^\dagger\mat S\mat M\}=\tr\{\mat m\mat m^\dagger\}=\tr\{\op\rho\}=1$.
Substituting $\lag$ by \eq{eq:lag} in \eq{eq:Mdot} leads to the final equation of motion for $\mat M$ [\eq{eq:vMCG-NOSSE-M}].

To derive the \gls{EOM} for $z_{\alpha k}$ we replace $\mat m^\dagger$ by its explicit form [\eq{eq:ansatz-m}] in \eq{eq:vMCG-NOSSE-lambdakalpha}
\bea\nonumber
&&\sum_{ab}M_{ba}^*\bra{\frac{\partial \varphi_b}{\partial  z_{\alpha k}}}\Big\{\sum_c\Big(\ket{\varphi_c}\dt{M}_{ca} +\ket{\dt{\varphi}_c}M_{ca}\\[-0.3cm]\nonumber
&& \hspace{2cm} + \lag \ket{\varphi_c}M_{ca}\Big) - \ket{\sqrtLiouv_a[\mat m]}\Big\}=0.\\
\eea
Substituting ${\dot M}_{ca}$ by \eq{eq:Mdot} gives
\bea\nonumber
&&\sum_{a}M_{ka}^*\bra{\frac{\partial \varphi_k}{\partial  z_{\alpha k}}}\Big\{\sum_c\Big(\ket{\varphi_c}[\mat S^{-1}\mat K - \mat S^{-1}\mat\tau\mat M - \lag \mat M]_{ca}\\[-0.3cm]\nonumber
&& \hspace{2cm} +\ket{\dt{\varphi}_c}M_{ca} + \lag \ket{\varphi_c}M_{ca}\Big) - \ket{\sqrtLiouv_a[\mat m]}\Big\}\\\nonumber
&&=\sum_{a}\bra{\frac{\partial \varphi_k}{\partial  z_{\alpha k}}}\Big\{\sum_c\big(\op 1 - \op P_{N_b}\big)\ket{\dt{\varphi}_c} M_{ca}\\[-0.3cm]\nonumber
&& \hspace{3cm} + (\op P_{N_b}-1) \ket{\sqrtLiouv_a[\mat m]}\Big\}M_{ka}^*\\
&&=0.
\eea
Expanding the time derivative using the chain rule in terms of the basis set 
parameters gives
\bea
&&\sum_{l\beta}\bra{\frac{\partial \varphi_k}{\partial z_{\alpha k}}}\Big[\op 1 - \op P_{N_b}\Big]\ket{\frac{\partial \varphi_l}{\partial z_{l\beta}}} [\mat M\mat M^\dagger]_{lk}\dot{ z}_{l\beta} \nonumber\\
&&\hspace{1.5cm}-\sum_l\bra{\frac{\partial \varphi_k}{\partial z_{\alpha k}}}\Big[\op 1 - \op P_{N_b}\Big]\ket{\sqrtLiouv_l[\mat m]}M_{kl}^* \nonumber\\
&&=\sum_{l\beta}C_{kl}^{\alpha\beta}\dot{ z}_{l\beta} - Y_k^\alpha =0,
\eea}
where we used the definitions from Eqs.~(\ref{eq:vMCG-NOSSE-C}) and (\ref{eq:vMCG-NOSSE-Y}). Adopting the matrix notation the last equation can be rewritten as \eq{eq:vMCG-NOSSE-lambda}.

%

\end{document}